\documentclass[12pt]{article}
\usepackage[english,activeacute]{babel}
\usepackage{natbib}
\usepackage{comment}
\usepackage{float}
\usepackage[hidelinks]{hyperref}
\usepackage{mathrsfs}
\usepackage{enumitem}
\usepackage[font={small,it}]{caption}
\usepackage{amsmath,amsfonts,amsthm,amssymb}
\usepackage{bm,rotating,multirow,dsfont,graphicx}
\usepackage[usenames, dvipsnames]{color}
\usepackage{url}
\usepackage{multicol}
\usepackage{multirow}
\usepackage[T1]{fontenc}
\usepackage{flafter}
\usepackage{appendix}
\usepackage{subfigure}
\usepackage{xcolor}
\usepackage{soul}
\usepackage[all]{xy} 
\usepackage{algorithm}
\usepackage{algpseudocode}
\makeatletter
\def\hlinewd#1{%
	\noalign{\ifnum0=`}\fi\hrule \@height #1 %
	\futurelet\reserved@a\@xhline}
\makeatother
\newcommand\iid{\mathrel{\overset{\makebox[0pt]{\mbox{\normalfont\tiny\sffamily iid}}}{\sim}}}

\newcommand\simind{\mathrel{\overset{\makebox[0pt]{\mbox{\normalfont\tiny\sffamily ind}}}{\sim}}}

\newcommand{\ubf}{\boldsymbol{u}}

\newcommand{\Ybf}{\mathbf{Y}}

\newcommand{\Ibf}{\mathbf{I}}

\newcommand{\thebf}{\boldsymbol{\theta}}

\newcommand{\Ip}{\mathcal{I}}
\newcommand{\Op}{\mathcal{O}}
\newcommand{\Real}{\mathbb{R}}

\newcommand{\Prob}{\textsf{Pr}}
\newcommand{\Esf}{\textsf{E}}
\newcommand{\CVsf}{\textsf{CV}}
\newcommand{\covsf}{\textsf{Cov}}
\newcommand{\tr}{\textsf{T}}

\newcommand{\psf}{\textsf{p}}

\newcommand{\expsf}{\textsf{exp}}
\newcommand{\expitsf}{\textsf{expit}}
\newcommand{\logitsf}{\textsf{logit}}

\newcommand{\Normalsf}{\textsf{N}}

\newcommand{\GammaIsf}{\textsf{GI}}
\newcommand{\Exponencialsf}{\textsf{Exp}}
\newcommand{\Bersf}{\textsf{Ber}}
\newcommand{\Betasf}{\textsf{Beta}}

\newcommand{\mathnorm}[1]{\left|\left| #1 \right|\right|}

\def\@roman#1{\romannumeral #1}

\begin{document}

\def\spacingset#1{\renewcommand{\baselinestretch}{#1}\small\normalsize}\spacingset{1}

\title{Influence Networks: \\ Bayesian Modeling and Diffusion}

\date{}

\author{
    Samuel Sánchez-Gutiérrez, Universidad Nacional de Colombia \\
    Juan Sosa, Universidad Nacional de Colombia, Colombia\footnote{Corresponding author: jcsosam@unal.edu.co.} \\
    Carolina Luque, Universidad Ean, Colombia
}

\maketitle

\begin{abstract} 
\noindent 
In this article, we make an innovative adaptation of a Bayesian latent space model based on projections in a novel way to analyze influence networks. 
By appropriately reparameterizing the model, we establish a formal metric for quantifying each individual's influencing capacity and estimating their latent position embedded in a social space.
This modeling approach introduces a novel mechanism for fully characterizing the diffusion of an idea based on the estimated latent characteristics.
It assumes that each individual takes the following states: Unknown, undecided, supporting, or rejecting an idea.
This approach is demonstrated using a influence network from Twitter (now $\mathbb{X}$) related to the 2022 Tax Reform in Colombia. 
An exhaustive simulation exercise is also performed to evaluate the proposed diffusion process.
\end{abstract}


\spacingset{1.1} 

\section{Introduction}\label{sec_intro}

The study of information arising from the interconnection between elements in a system is key to understand many phenomena. 
The structure formed by these elements (individuals or actors) and their interactions (ties or connections) is commonly known as graph, social network, or simply a network. 
Examples of networks are prevalent across various research areas, including Finance (studying alliances and conflicts between countries as part of the global economy), Social Sciences (studying interpersonal social relationships and social collaboration schemes, such as legislative co-sponsorship networks), Biology (studying interaction arrangements among genes, proteins, or organisms), Epidemiology (studying the spread of an infectious disease), and Computer Science (studying the Internet, the World Wide Web, and communication networks), among many others. 
A few examples illustrate that both the entities and the connections in networks are varied and diverse, ranging from individuals to organizations and from friendship to communication, respectively \citep{sosa2021review}.

For further exploration of social network analysis, including basic definitions and properties along with various applications across multiple scenarios, recommended sources include \cite{wasserman1994social}, \cite{scott2011social}, \cite{newman2003structure}, \cite{kolaczyk2014statistical}, and \cite{menczer2020first}. 
Additionally, specialized reviews are available that address specific topics related to social networks.
For instance, \cite{sosa2021review} offers a comprehensive review of latent space models for social networks (\citealt{hoff2002latent}, \citealt{hoff2007modeling}, \citealt{hoff2009multiplicative}), while \cite{zhang2016dynamics} provides a valuable review covering a range of topics from dynamic models for information diffusion (studying actors and connections within a network over time) in order to identify leaders in the system. 
The literature in the various fields of social networks is extensive. 
For example, from the dynamic perspective of social network analysis, there is a wide variety of proposals for modeling the evolution of the system over time, including works by \cite{durante2014nonparametric}, \cite{hoff2015multilinear}, \cite{sewell2015latent}, \cite{sewell2016latent}, \cite{sewell2017latent}, \cite{gupta2019generative}, \cite{kim2018review}, \cite{turnbull2020advancements}, \cite{betancourt2020modelling}, among many others.

Currently, online social networks (e.g., Facebook, Instagram, TikTok, among others) represent one of the primary means of information dissemination and, therefore, they have a significant impact on the formation of public opinion. 
This work proposes using graph theory and Bayesian modeling techniques to abstract user interactions in online social networks and comprehensively characterize phenomena associated with influence. 
Specifically, we provide a full Bayesian model to estimate the social position of users in order to explain how they relate to each other. 
The model give us the means to quantify the users' ability to influence their neighbors in the system to adopt a specific stance on a relevant topic.
Then, we develop a dynamic model from the former model parameters to fully characterize how an idea spreads through out the system.

In this way, we extensively study the parameters of the explanatory interaction model, and subsequently, we explore several synthetic scenarios to investigate the diffusion of an idea.
Even though a common approach to modeling the diffusion of an idea is through compartmental models widely used in epidemiology (e.g., SIS, SIR, SEIR, among others), it is well known that typically networks through which actors interact are not homogeneous, meaning that individuals are not homogeneously mixed (\citealt{linkletter2007spatial}, \citealt{barrat2008dynamical}, \citealt{kolaczyk2014statistical}).
Therefore, it is imperative to develop an alternative to characterize completely, accounting for such heterogeneous relational patterns (typically observed in online social networks).

There are multiple alternatives in the literature regarding measuring influencing capacity of individuals. 
\citet[Chapter 8.2]{zafarani2014social} provides an excellent introduction to the topic, covering everything from threshold models to more advanced proposals such as that given in \cite{yang2010modeling}, which quantifies individuals' influence by means of a time-oriented curve. 
Other successful proposals, such as those provided in \cite{romero2011influence} and \cite{bakshy2011everyone}, quantify influencing capacity using scalar quantities. 
We also opt for a scalar representation due to its implicit parsimony.  
However, unlike the previous alternatives, we do so using a probabilistic model that allows us to quantify the uncertainty associated with the phenomenon of interest. 
Specifically, we use a novel adaptation of the latent space projection model of  \cite{hoff2002latent} as a proxy to to estimate the influencing capacity. 
Nonetheless one of the most common applications of projection models is modeling international trade networks (e.g., \citealt{hoff2004modeling}, \citealt{ward2007persistent}), these models can also be applied to a wide range of directed networks such as online social networks.

This article is structured as follows: Section \ref{sec:influence-model} presents the characteristic aspects of an influence network. 
Section \ref{sec:difusion} proposes a novel diffusion model based on the social positioning of actors. 
Section \ref{sec:reald} provides an application in the context of an influence network on Twitter (now $\mathbb{X}$) related to tweets, retweets, quotes, and comments published in Spanish during the 32 hours preceding the approval of the Tax Reform on November 4, 2022 (Law 2277 of 2022) by the Congress of the Republic of Colombia. 
Section \ref{sec:difsimulation} conducts an exhaustive simulation exercise to study the proposed diffusion model. 
Finally, Section \ref{sec:discusion} presents the findings obtained and future research alternatives.

\section{Modeling a influence network}\label{sec:influence-model}

Let $\mathbf{Y} = [y_{i,j}]$ be the adjacency matrix associated with a \textit{directed binary network}, such that $y_{i,j} = 1$ if there is a directed link from vertex $i$ to vertex $j$, and $y_{i,j} = 0$ otherwise, for $i, j = 1, \ldots, N$ with $i \neq j$.
In order to model the connections given in the network, we assume that $\mathbf{Y}$ is an exchangeable array of rows and columns (i.e., $\Prob\left(\mathbf{Y} = [y_{i,j}]\right) = \Prob\left(\mathbf{Y} = [y_{\pi(i),\pi(j)}]\right)$ for every permutation $\pi$ of integers $1, \ldots, N$). Thus, we consider the a \textit{latent space projection model} \citep{hoff2002latent}, given by 
\begin{equation}\label{eq:likelihoodhoff}
y_{i,j} \mid \Op_i, \ubf_i, \ubf_j \simind \Bersf \left[\expitsf \left( \Op_i + \frac{\ubf_i\cdot \ubf_j}{\mathnorm{\ubf_i}} \right) \right]\,, 
\end{equation} 
where $\expitsf(x) = 1/\left(1+\exp(-x)\right)$ is the inverse function of the logit function $\logitsf(x) = \log(x/(1-x))$, $\Op_i$ is a random effect representing the propensity of observing a directed link from vertex $i$, $\ubf_i = (u_{i,1}, \ldots, u_{i,p})$ is a random vector in $\Real^{p}$ representing the latent (unobserved) characteristics of vertex $i$ in the social space where the individuals of the system are embedded, $\boldsymbol{u}_i\cdot\boldsymbol{u}_j = \boldsymbol{u}_i^\tr\boldsymbol{u}_j = \sum_{k=1}^p u_{i,k}\,u_{j,k}$ is the dot product between $\boldsymbol{u}_i$ and $\boldsymbol{u}_j$, and finally, $\mathnorm{\boldsymbol{u}_i} = \sqrt{\boldsymbol{u}_i \cdot \boldsymbol{u}_i}$ is the Euclidean norm of $\boldsymbol{u}_i$, assuming that the latent dimension $p$ is a known fixed quantity. 
For example, $p=2$ is a reasonable choice that simplifies the visualization and description of social relationships.

In the context of influence networks, we call \textit{positions} (preferences or ideas) to those latent characteristics that generate the social space. 
Thus, under the model specification \eqref{eq:likelihoodhoff}, the probability of observing a directed link from one actor to another increases or decreases depending on the orientation of the corresponding positions.
Specifically, the vectors $\ubf_1,\ldots,\ubf_N$ allow us to identify and quantity the affinity or dissimilarity between individuals' preferences, given that $\ubf_i\cdot\ubf_j < 0$, $\ubf_i\cdot\ubf_j=0$ and $\ubf_i\cdot\ubf_j>0$, depending on whether the angle $0\leq\theta_{i,j}\leq\pi$ between $\ubf_i$ and $\ubf_j$ is obtuse, right, or acute, respectively (as long as $\ubf_i$ and $\ubf_j$ are not $\boldsymbol{0}_p$).

When the network links represent an influence relationship, in order to quantify the ability to influence others, the degree of influenceability, and the similarity between the opinion states of individuals in the system, we propose the following reparameterization of the projection model 
\begin{equation}\label{eq:repar} 
\Ip_j = \mathnorm{\ubf_j}
\qquad\text{and}\qquad 
\tau_{i,j} = 
\left\{\ \begin{array}{ll} \cos\theta_{i,j} = \frac{\ubf_i\cdot \ubf_j}{\mathnorm{ \ubf_i }\,\mathnorm{\ubf_j}}, & \text{if } \ubf_i,\ubf_j \neq \boldsymbol{0}_p,\\ 0, & \text{otherwise,} \end{array} \right. 
\end{equation} 
where $\Ip_j \geq 0$ for all $j=1,\ldots,N$, and $-1 \leq \tau_{i,j} \leq 1$ for all $i,j=1,\ldots,N$, with $i\neq j$. According to this reparameterization, the model \eqref{eq:likelihoodhoff} can be written as, 
$$ y_{i,j} \mid \Op_i, \Ip_j, \tau_{i,j} \simind \Bersf \left[\expitsf\left( \Op_i + \tau_{i,j} \,\Ip_j \right) \right]\,, $$ 
which we refer to as \textit{influence model}.
According to this parameterization, $\tau_{i,j}$ can be though as a measure of similarity in the social space, whereas $\Ip_j$ serves as a weighting factor of the individual being influenced, modulating the similarity between stances:
\begin{itemize}
\item If the stances of individuals $i$ and $j$ do not align with each other, then $\tau_{i,j}\longrightarrow 0$ and consequently $\Ip_j$ does not affect the probability that $i$ influences $j$. 
\item If the stances of individuals $i$ and $j$ do align with each other in the same direction, then $\tau_{i,j}\longrightarrow 1$ and consequently $\Ip_j$ has a positive effect on the probability that $i$ influences $j$.
\item If the stances of individuals $i$ and $j$ do align with each other in opposite directions, then $\tau_{i,j}\longrightarrow -1$ and consequently $\Ip_j$ has a negative effect on the probability that $i$ influences $j$.
\end{itemize}

From the previous statements, it follows that if $\Ip_j$ takes ``large'' values, then the individual $j$ tends to take extreme positions by following those who share their ideas and ignoring the influence of others who do not. 
On the other hand, if $\Ip_j$ takes ``small'' values, then $j$ is likely to be a moderate individual opened to the influences of other individuals equally. 
Finally, the effect of $\Op_i$ on the probability that an individual $i$ influences an individual $j$ does not depend on the affinity between the individuals' ideas but rather on other factors, such as the sociability of individual $i$ in the social network.
From our previous discussion, we expect that $\Op_i$ turns out to be positively correlated with the out-degree of individual $i$.

\subsection{Computation}\label{sec:posterior}

The Bayesian inference paradigm (e.g., \citealt{hoff2009first}, \citealt{gelman2014bayesian}, \citealt{reich2019bayesian}) allow us to estimate the parameters of a model in high dimensions, particularly in cases such as the one given in model \eqref{eq:likelihoodhoff}, where the number of parameters grows quadratically as the number of individuals in the social system increases. 
Such estimation capability arises because the prior distribution naturally introduces specific mechanisms of identifiability and regularization into the estimation process.

\begin{figure}[!htb]
     \centering
     \resizebox{.49\textwidth}{!}{\input{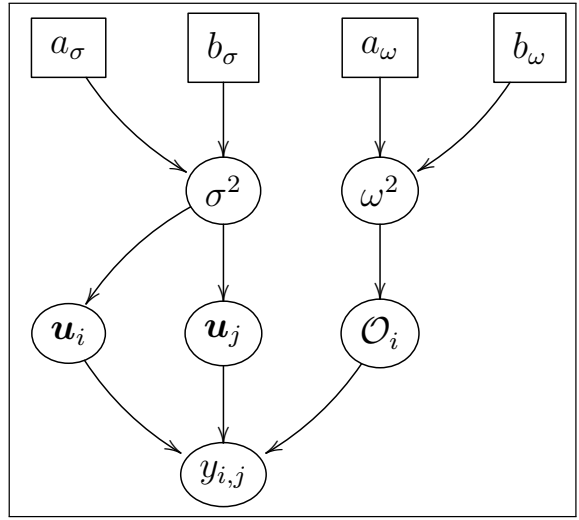}}
     \caption{DAG for the influence model.}
     \label{fig:dag}
\end{figure}

In order to perform a fully Bayesian analysis of the influence network according to model \eqref{eq:likelihoodhoff}, it is necessary to specify a prior distribution for the baseline influence probabilities $\Op_1,\ldots,\Op_N$ along with the latent characteristics $\ubf_1,\ldots,\ubf_N$.
A standard choice consists in setting a hierarchical prior distribution of the form
\begin{equation*}\label{eq:prior}
    \Op_i  \mid  \omega^2 \iid \textsf{N}(0,\omega^2)\,,
    \qquad \omega^2 \sim \textsf{GI}(a_\omega,b_\omega)\,,
    \qquad u_{i,k}  \mid  \sigma^2 \iid \textsf{N}(0,\sigma^2)\,, 
    \qquad \sigma^2 \sim \textsf{GI}(a_\sigma,b_\sigma)\,,
\end{equation*}
for $i=1,\ldots,N$ and $k=1,\ldots,p$, where $a_\omega>0$, $b_\omega>0$, $a_\sigma>0$, and $b_\sigma>0$ are fixed known quantities, and $\textsf{N}(\mu,\sigma^2)$ and $\textsf{GI}(\alpha,\beta)$ denote the Normal and Inverse Gamma distributions, respectively. 
Therefore, our influence model has $N(p+1)$ unknown parameters to estimate, namely, $\mathbf{\Upsilon} = (\Op_1,\ldots,\Op_N, \ubf_1,\ldots,\ubf_N)$, associated with four hyperparameters $a_\omega, b_\omega, a_\sigma, b_\sigma$, which must be selected appropriately to ensure adequate model performance.
Figure \ref{fig:dag} shows the representation of the model using a directed acyclic graph (DAG).

For a specific latent dimension $p$ (e.g., $p=2$), the posterior distribution $p(\mathbf{\Upsilon}\mid \mathbf{Y})$ can be empirically approximated using Markov chain Monte Carlo (MCMC; e.g., \citealt{hoff2009first}, \citealt{gelman2014bayesian}, \citealt{reich2019bayesian}) algorithms.
Specifically, we develop a Gibbs sampler with semi-conjugate and Metropolis-Hastings steps \citep{robert2010introducing} to obtain dependent but approximately independent and identically distributed samples $\mathbf{\Upsilon}^{1},\ldots,\mathbf{\Upsilon}^{B}$ from $p(\mathbf{\Upsilon}\mid \mathbf{Y})$.
In Appendix A we present the complete conditional distributions of each model parameter, along with the MCMC algorithm to generate samples from the posterior distribution. All the code necessary to reproduce our findings is available at \url{https://github.com/Samuel-col/InfluenceNetworks}.

\section{Diffusion model} \label{sec:difusion}

The \textit{diffusion of an idea} or \textit{cascade} (e.g., \citealt{watts2002identity}, \citealt{watts2007influentials}, \citealt{zhang2016dynamics}) is the process in which the members of a social system transmit information to maintain or change their state (cognitive position regarding a specific topic). 
Thus, for a \textit{public issue} (a subject susceptible to diffusion), we assume that each individual can adopt one of the following states: Unknown (I), Undecided (U), Support (S), or Reject (R).

\begin{figure}[!htb]
    \centering
    \resizebox{.6\textwidth}{!}{\input{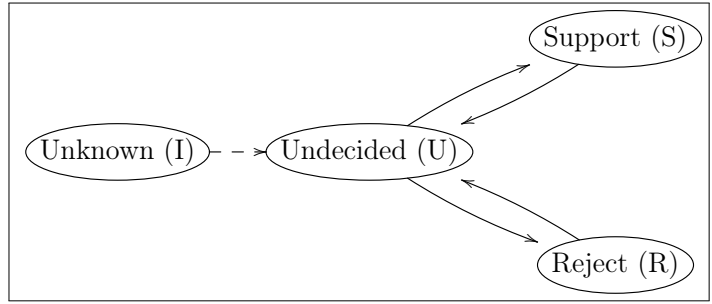}}
    \caption{Change of states diagram in a diffusion process.}
    \label{fig:estados}
\end{figure}

Before a diffusion process begins, typically most individuals are in state I. 
When the process begins, individuals in states U, S, or R \textit{inform} their neighbors (adjacent vertices in the influence network) in state I, who in turn change (mutate, transition, or jump) to state U (an individual in state U cannot revert to state I). 
Additionally, individuals in state R \textit{influence} their neighbors to change from state S to state U and from state U to state R (similarly, individuals in state S promote comparable changes). 
Thus, in a diffusion process, individuals are exposed to changes through information or influence, denoted in Figure \ref{fig:estados} with $\dashrightarrow$ and $\longrightarrow$, respectively.

Typically, when a diffusion process begins, the individuals' ability to inform and influence others decreases over time.
That is, the probability that an individual convinces his neighbors to change their state decreases after adopting their current stance.
Therefore, it is quite reasonable to consider the time associated with state changes as random variables governed by a distribution with some particular decay.

For this purpose, we define $\kappa_{i,j}$ as the (remaining) time for individual $i$ to cause a change by means of information in their neighbor $j$ (i.e., when $i$ is in state U, R, or S and $j$ is in state I). 
Similarly, $\rho_{i,j}$ is defined as the (remaining) time for individual $i$ to cause a change through influence in their neighbor $j$ (i.e., when $i$ is in state R or S and $j$ is in state U, R, or S).
Thus, we assume that $\kappa_{i,j}$ and $\rho_{i,j}$ are random variables such that
$$
\kappa_{i,j}\mid\Op_i \simind \Exponencialsf\left(e^{\Op_i}\right)
\qquad\text{y}\qquad
\rho_{i,j}\mid\Op_i,\Ip_j,\tau_{i,j} \simind \Exponencialsf\left(e^{\Op_i + \tau_{i,j} \Ip_j}\right)\,,
$$
where $\Exponencialsf(\lambda)$ denotes the Exponential distribution with mean $1/\lambda$. 
Note that the distribution of times explicitly depends on the parameters of the proposed model in section \ref{sec:influence-model} to characterize the influence network. 
This articulation is fundamental for linking directly influence relationships with diffusion processes. 
For example, under this formulation, it follows that the relative possibility of $i$ influencing $j$ matches the expected time for $i$ to cause a change through influence in $j$, i.e.,
$$
\logitsf \,\Prob(y_{i,j} = 1 \mid \Op_i,\ubf_i,\ubf_j) = e^{-(\Op_i + \tau_{i,j} \Ip_j)} = \textsf{E}(\rho_{i,j} \mid \Op_i,\Ip_j,\tau_{i,j})\,,
$$
which is particularly appealing for fully characterizing influence from a relational and a cognitive perspective.

In computational terms, the diffusion process is carried out sequentially through \textit{individual} changes until a predetermined stopping criterion is reached.
Each update in the state structure involves only one individual at a time, as the probability of two individuals changing state simultaneously is 0. 
Then, in order to determine a state change, all possible transitions are considered based on the current state structure and the links given in the influence network.
After that, the corresponding time to cause a change, whether through information or influence, is generated for each possibility.
In this way, the following state change corresponds to the individual associated with the smallest of all the times generated previously.
This criterion for updating the state structure ensures that the random mechanism associated with the transition times is preserved. 
Such a a fact is guaranteed by the memoryless property of the Exponential distribution.

The algorithm stops when there are no more transitions to perform or when the total number of individuals in each of the four states does not vary by more than $5\%\,N$ after $3N$ changes, where $N$ is the number of individuals in the system. 
Typically, diffusion processes require approximately $2N$ transitions, since each individual only needs two changes to reach states S or R. 
For this reason, using $3N$ changes as stopping criterion is perfectly reasonable for terminating the proposed algorithm.
In Appendix C, we provide the pseudocode for simulating our diffusion process in a influence network based on the algorithm we just described.

\section{Case study: A Twitter influence network}\label{sec:reald}

In this section, we study a influence network from Twitter (now $\mathbb{X}$).
Using \texttt{Hoaxy} \citep{hoaxyIU}, we collected a dataset including tweets, retweets, quotes, and comments published in Spanish about the Tax Reform in Colombia, during the 32 hours leading up to its approval by the Colombian Congress on November 4, 2022 (Law 2277 of 2022).
In the sampling design, users are stratified according to their level of popularity to account for both popular users and regular users of the social network, which allows for capturing perspectives from both popular and ordinary users on the social network.
This approach can offer a more comprehensive and balanced perspective on social media reactions to the Tax Reform.

\begin{figure}[!htb]
    \centering
    \includegraphics[width = 1.0\textwidth]{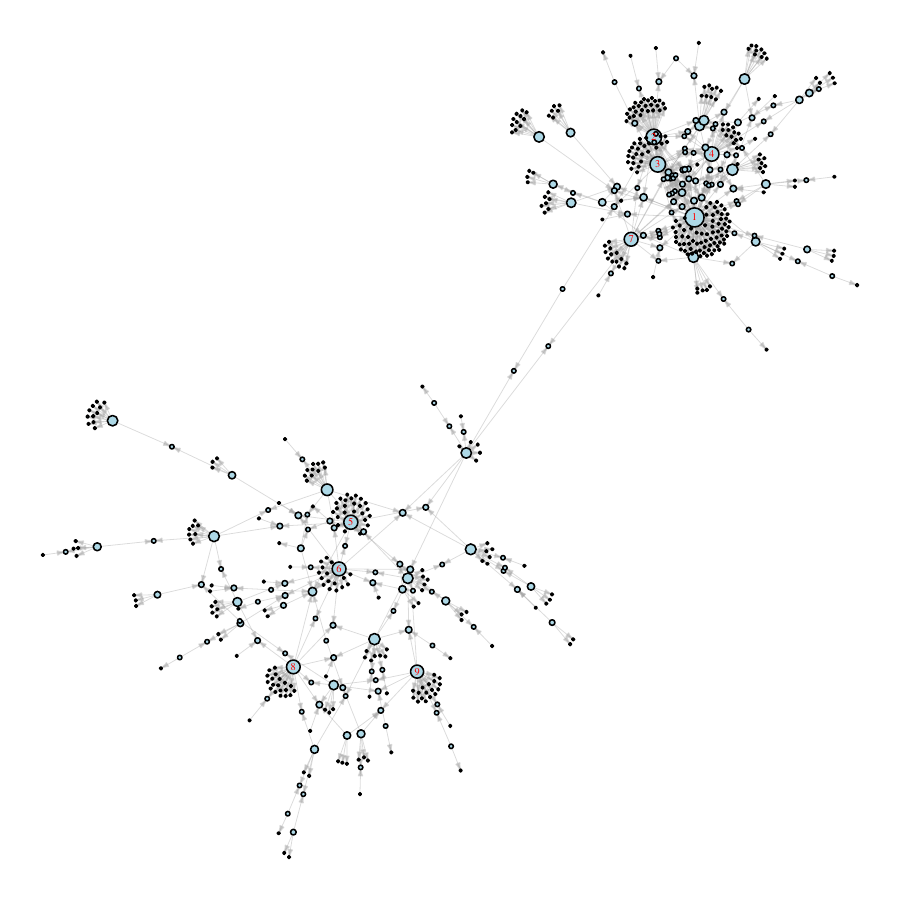}
    \caption{Twitter influence network regarding the Colombian Tax Reform in Colombia. The node size is proportional to out-degree.}
    \label{fig:rt-red}
\end{figure}

Here, we establish influence relationships through retweets. 
If user B shares a tweet posted by user A on his profile without adding any additional comments, then we say that an relationship of influence exists from A to B.
Thus, dyads associated with influence relationships are defined as $y_{i,j} = 1$ if $j$ retweets $i$, and $y_{i,j} = 0$ otherwise.
The influence network under study corresponds to the giant component (the largest connected subgraph) of all observed influence relationships.
This network consists of $N = 634$ actors and $745$ connections.

Table \ref{tab:rt-populares} lists the individuals with the highest out-degree (i.e., the most retweeted users).
The four most retweeted individuals belong to the group that supports the Tax Reform. 
Figure \ref{fig:rt-red} presents a visualization of the influence network.
Users are clearly divided into two groups, with a few links between them.
Additionally, the most popular individuals appear across both groups. 
This structural behavior of the network indicates a phenomenon of \textit{ideological polarization} concerning the reform.
Each group also displays similar generative structures typical of a preferential attachment model (the ``rich get richer'' effect; \citealt{barabasi1999emergence}).
Both the complete network and each group exhibit scale-free properties (the probability that a vertex connects to $k$ vertices is proportional to $k^{-\gamma}$, with $\gamma > 0$; \citealt{newman2003structure}).

\begin{table}[!htb]
\footnotesize
\centering
\setlength{\tabcolsep}{2pt} 
\begin{tabular}{ccll}
\hline
\textbf{ID} & \textbf{Retweets} & \multicolumn{1}{c}{\textbf{User}} & \multicolumn{1}{c}{\textbf{Description}}                                                                                                                   \\ \hline
1  & 108     & petrogustavo   & President of Colombia and promoter of the Reform                                                                                                         \\
2  & 45      & GustavoBolivar & 
Ex-Senator                                                                                                                              \\
3  & 44      & wilsonariasc   & Senator                                                                                                                                 \\
4  & 34      & IvanCepedaCast & Senator                                                                                                                               \\
5  & 31      & MiguelUribeT   & Senator                                                                                                                                \\
6  & 29      & alertaLatam    & Account for spreading false political news \citep{lasillavacia2023alertalatam} \\
7  & 29      & luiscrh        & Director DIAN (National Tax and Customs Directorate)                                                                                      \\
8  & 28      & jarizabaletaf  & Lawyer and columnist without any public office positions                                                                                       \\
9  & 24      & CeDemocratico  & Account of the political party opposing Gustavo Petro's government                                                                               \\
10 & 16      & JERobledo      & Ex-senator                                                                                                                            \\ \hline
\end{tabular}
    \caption{Top 10 most retweeted actors in the Twitter influence network regarding to the Tax Reform in Colombia.}
    \label{tab:rt-populares}
\end{table}

\begin{table}[!b]
\footnotesize
\centering
\setlength{\tabcolsep}{2.5pt} 
    \begin{tabular}{lc}\hline
        \textbf{Measure}             &  \textbf{Value} \\ \hline
        Density                      &  0.001856 \\
        Transitivity                 &  0.000253 \\
        Assortativity                & -0.028710 \\
        Average distance             &  6.959768 \\
        Average degree               &  2.350158 \\
        Standard deviation of degree &  5.854511 \\
        Clustering modularity        &  0.783321 \\ \hline
    \end{tabular}
    \caption{Descriptive topological statistics of the influence network regarding the Tax Reform in Colombia.}
    \label{tab:rt-stats}
\end{table}

Table \ref{tab:rt-stats} presents several descriptive topological measures of the influence network. 
The network exhibits low connectivity, showing only $0.18\%$ of all possible influence relationships.
On average, each individual connects with $2.35$ of the other $633$ users in the system, either as an influencer or as someone being influenced. 
Additionally, the degree variability indicates a wide range of variation regarding the influence capacities. 
The corresponding coefficient of variation is $249\%$.
Finally, the network's transitivity is notably low, with the fraction of transitive triplets (triangles) being just $0.02\%$. 
Finding transitive influence relationships is uncommon because these relationships follow a hierarchical configuration, i.e., relationships typically flow from more influential actors to less influential ones, which prevents the formation of triangles.

\subsection{Modeling}

We fit the influence model from Section \ref{sec:influence-model} with $p=2$ latent dimensions and three different hyperparameter settings (Table \ref{tab:hypersettings}), using the MCMC algorithm described in Appendix B with $B=5000$ samples from the posterior distribution after a warm-up period of $5000$ iterations and a systematic sampling of every $10$ observations. 
None of the chains exhibit issues with convergence to the stationary distribution. 
The initial conditions correspond to diffuse prior information (i.e., non-informative) and do not favor any particular region of the parameter space, given the high prior variability conditions.
Fitting the model under different prior conditions is a common practice in the Bayesian paradigm, since it allow us to assess the sensitivity of the posterior inference to the prior distribution.

Table \ref{tab:hypersettings} presents the prior settings, execution time (minutes per iteration), and the Deviance Information Criterion (DIC; \citealt{spiegelhalter2002bayesian}, \citealt{spiegelhalter2014deviance}, \citealt{gelman2014bayesian}) for each case.
All execution times are approximately equal (0.02 minutes per iteration). 
Similarly, the DIC values are very similar across all the scenarios. 
Moreover, the posterior means of the latent features $\boldsymbol{u}_1,\ldots,\boldsymbol{u}_N$ under the three configurations are similar after performing rotations and reflections of the social space (not shown here). 
Previous studies indicate that the likelihood of the projection model remains invariant to these transformations \citep{hoff2002latent}. 
Therefore, no computational, predictive, or inferential reasons favor one hyperparameter configuration over another.

\begin{table}[!htb]
\footnotesize
\centering
\setlength{\tabcolsep}{2.5pt} 
    \begin{tabular}{ccccccccccc}\hline
        \textbf{Configuration} & $a_\omega$ & $b_\omega$ & $\Esf(\omega^2)$ & $\CVsf(\omega^2)$ & $a_\sigma$ & $b_\sigma$ & $\Esf(\sigma^2)$ & $\CVsf(\sigma^2)$ & \textbf{Time} & \textbf{DIC}\\\hline
        1 & 1 & 1 & $\infty$ & $\infty$ & 1 & 1 & $\infty$ & $\infty$ & 0.021 & 9079.079\\
        2 & 2 & 1 & 1 & $\infty$ & 2 & 1 & 1 & $\infty$ & 0.020 & 9108.588\\
        3 & 3 & 2 & 1 & 1 & 3 & 2 & 1 &1 & 0.024 & 8979.182\\\hline
    \end{tabular}
    \caption{Prior settings, execution time (minutes per iteration), and Deviance Information Criterion (DIC). Processing was done on a laptop with 8 GB of RAM and a 2.4 GHz Intel Pentium processor.}
    \label{tab:hypersettings}
\end{table}

\subsection{Posterior inference}

The posterior inference is conducted with $p = 2$ latent dimensions and using the first prior configuration from Table \ref{tab:hypersettings}.
This case represents the scenario with the highest prior uncertainty (least amount of information) since $\textsf{CV}(\omega^2) = \textsf{CV}(\sigma^2) = \infty$. 
The posterior inference relies on $B=5000$ samples from the posterior distribution, following the model fitting guidelines described previously.

\begin{figure}[!hbt]
    \centering
    \subfigure[Twitter influence network.]{\includegraphics[width = 0.49\textwidth]{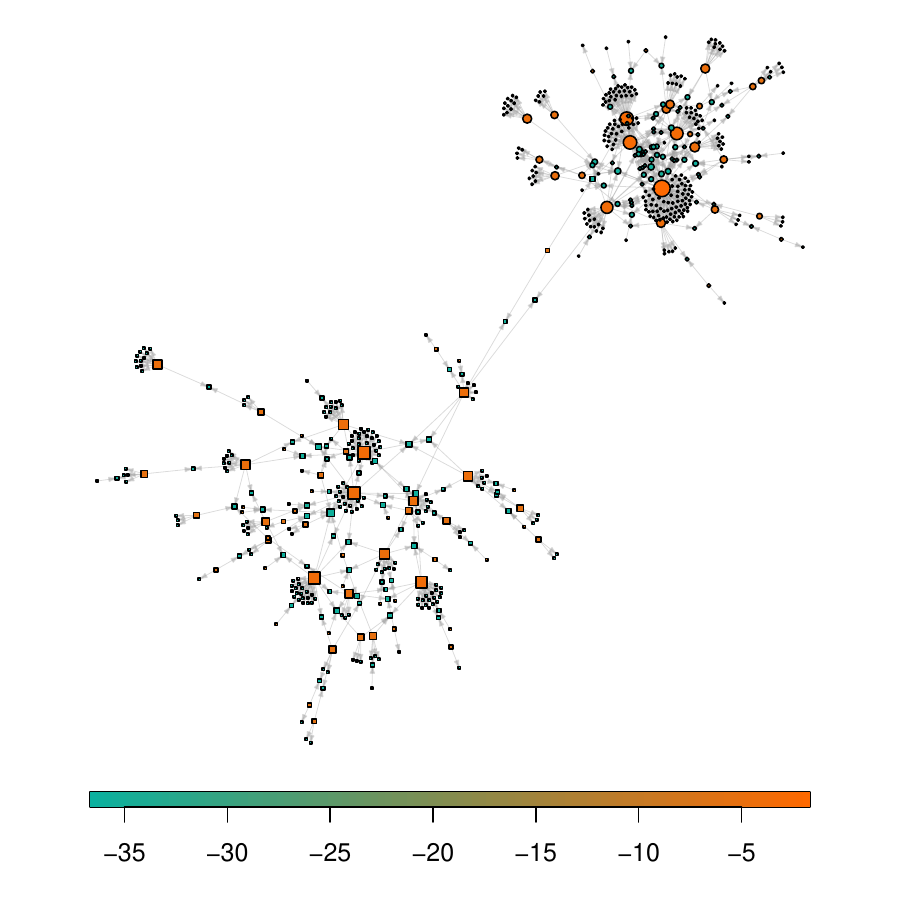}}
    \subfigure[Latent space.]{\includegraphics[width = 0.49\textwidth]{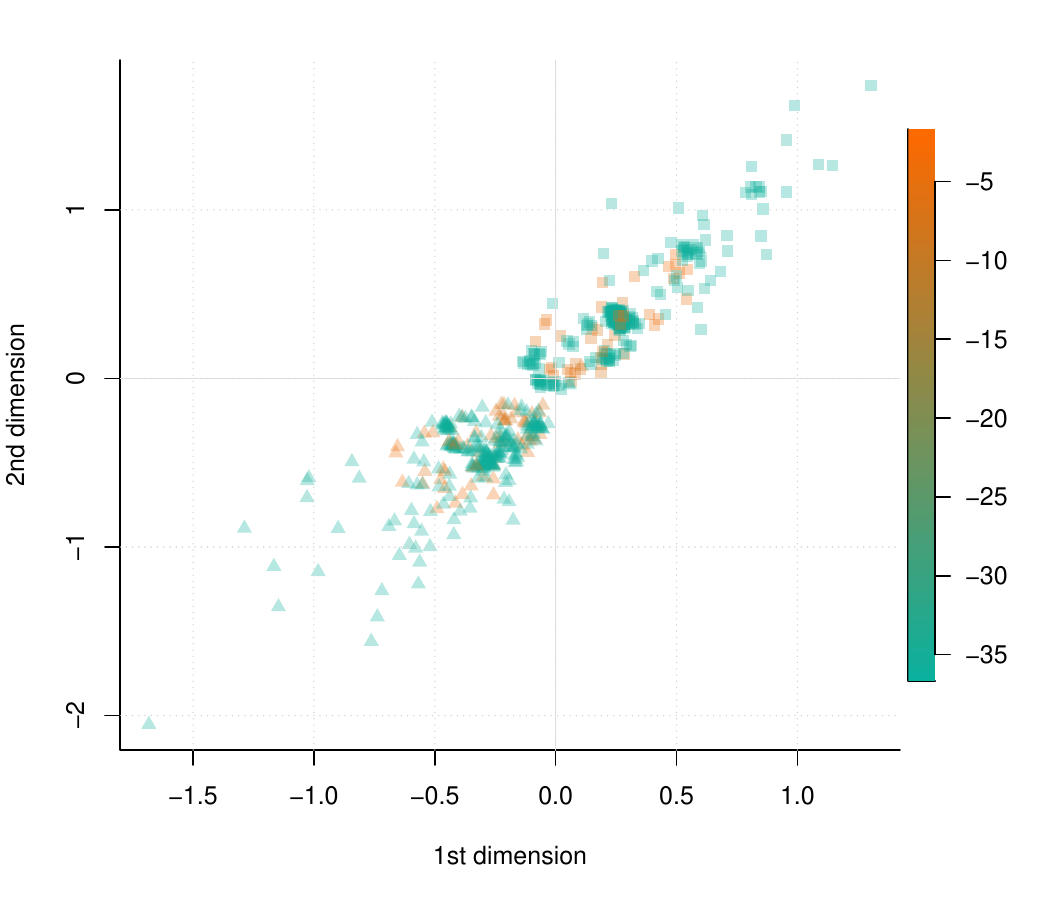}}
    \caption{Panel (a): Social influence network of Figure \ref{fig:rt-red}. Panel (b): Estimated latent features. Colors represent the estimated influencing capacity of each user, and shapes represent the communities built from the latent features.}
    \label{fig:rt-o-color}
\end{figure}

Table \ref{tab:rt-best-o} displays the top 10 users with the highest influencing capacity. 
This characteristic is quantified by the posterior mean of the global interaction propensity $\Esf(\Op_i \mid \Ybf) \approx \frac{1}{B}\sum_{b=1}^B \Op_i^{(b)}$.
The top 10 users in the ranking correspond to the most retweeted users (Table \ref{tab:rt-populares}).
The influencing capacities are organized into two groups: One with values less than $-30$ and another with values greater than $-10$ (left panel of Figure \ref{fig:rt-o-color}). 
No individual in the system has an influencing capacity between $-30$ and $-10$. 
Figure \ref{fig:rt-o-color} simultaneously shows the posterior means of $\boldsymbol{u}_1,\ldots,\boldsymbol{u}_N$ and $\Op_1,\ldots,\Op_N$, and the segmentation of users using FGA (\textit{Fast-greedy algorithm}; \citealt{newman2004fast}).
Users with higher influencing capacity (orange) have higher out-degrees (most retweeted individuals in the network). Additionally, users with lower influencing capacity (green) tend to retweet mainly one or very few influential users and rarely retweet non-influential users.

\begin{table}[!htb]
\footnotesize
\centering
\setlength{\tabcolsep}{2.5pt} 
\begin{tabular}{clcc}
  \hline
  ID & User & $\Esf(\Op_i \mid \Ybf)$ & $d^{\text{out}}_i$ \\ 
  \hline
  1  & petrogustavo   & -1.85 & 108 \\ 
  2  & GustavoBolivar & -2.94 & 45  \\ 
  3  & wilsonariasc   & -2.97 & 44  \\ 
  4  & IvanCepedaCast & -3.27 & 34  \\ 
  5  & MiguelUribeT   & -3.40 & 31  \\ 
  6  & luiscrh        & -3.44 & 29  \\ 
  7  & alertaLatam    & -3.48 & 29  \\ 
  8  & jarizabaletaf  & -3.52 & 28  \\ 
  9  & CeDemocratico  & -3.69 & 24  \\ 
  10 & JERobledo      & -4.12 & 16  \\ 
   \hline
\end{tabular}
    \caption{Top 10 users with the highest influencing capacity. 
    Out-degree $d^{\text{out}}_i$ and posterior mean of the global interaction propensity $\Esf(\Op_i \mid \Ybf)$ for the users.}
    \label{tab:rt-best-o}
\end{table}

The right panel of Figure \ref{fig:rt-o-color} shows that the two latent dimensions are highly correlated.
This finding underscores the efficiency of a single latent feature in simplifying the variability in the social space associated with the Tax Reform debate in Colombia.
Indeed, a principal component analysis confirms this finding, as the first component alone explains $96.7\%$ of the total variability.
This behavior is also evident in the two communities of users derived from the latent features using the FGA clustering algorithm.
A single latent feature explains $94.7\%$ and $84.4\%$ of the variability for the group that supports (circles) and the group that opposes (squares) the Tax Reform, respectively. 
Consequently, fitting the model using three or more latent dimensions is not necessary.

An important finding is the lack of significant evidence for a correlation between influencing capacity, $\Op$, and influencibility, $\Ip$.
To illustrate this, we constructed a credibility interval based on the $95\%$ confidence percentiles for $\rho_{\Op,\Ip} = \textsf{Cor}(\Op,\Ip)$. 
Such an interval is obtained by calculating the $2.5\%$ and $97.5\%$ percentiles of $\rho_{\Op,\Ip}^1, \ldots, \rho_{\Op,\Ip}^B$, where
\[
    \rho_{\Op,\Ip}^b = \frac{\sum_{i=1}^N \left(\Op_i^b - \bar{\Op}^b\right)\left(\Ip_i^b - \bar{\Ip}^b\right)}{\sqrt{\sum_{i=1}^N\left(\Op_i^b - \bar{\Op}^b\right)^2} \sqrt{\sum_{i=1}^N\left(\Ip_i^b - \bar{\Ip}^b\right)^2}}\,,
\]
with $\bar{\Op}^b = \frac{1}{N}\sum_{i=1}^N \Op_i^b$ and $\bar{\Ip}^b = \frac{1}{N}\sum_{i=1}^N \Ip_i^b$, for $b=1,\ldots,B$.
The resulting interval is $(-0.141, 0.012)$. 
Since this interval does not contain zero, we conclude that there is insufficient empirical evidence to claim that $\rho_{\Op,\Ip}$ significantly differs from $0$.

\begin{figure}[!htb]
    \centering
    \includegraphics[width = 0.6\textwidth]{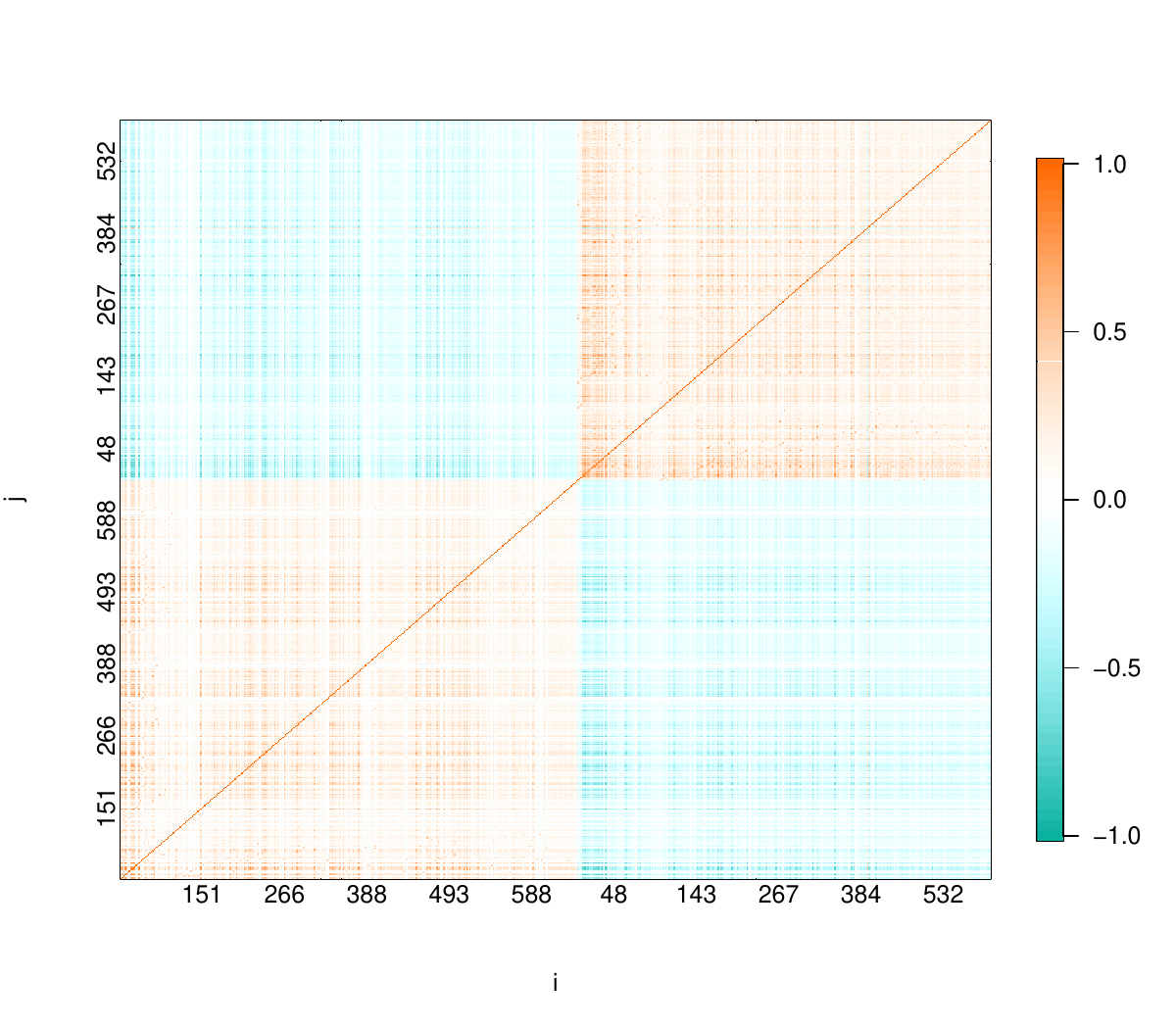}
    \caption{Estimated similarities $\tau_{i,j}$. Entries are arranged in rows and columns according to the communities identified from the latent features.}
    \label{fig:rt-tau}
\end{figure}

Figure \ref{fig:rt-tau} shows a pixel visualization of the estimated similarities $\Esf(\tau_{i,j} \mid \mathbf{Y})$. 
This a $N \times N$ matrix and the corresponding entries are arranged according to the two communities described earlier. 
Users do not exhibit completely opposing positions in the debate about the Tax Reform, since most of these similarities lie between $-0.5$ and $0.5$.
Additionally, it is important to note that the similarities align with the communities within the influence network, demonstrating that members of the same community tend to have similar positions.

\begin{figure}[!htb]
    \centering
    \includegraphics[width = 0.6\textwidth]{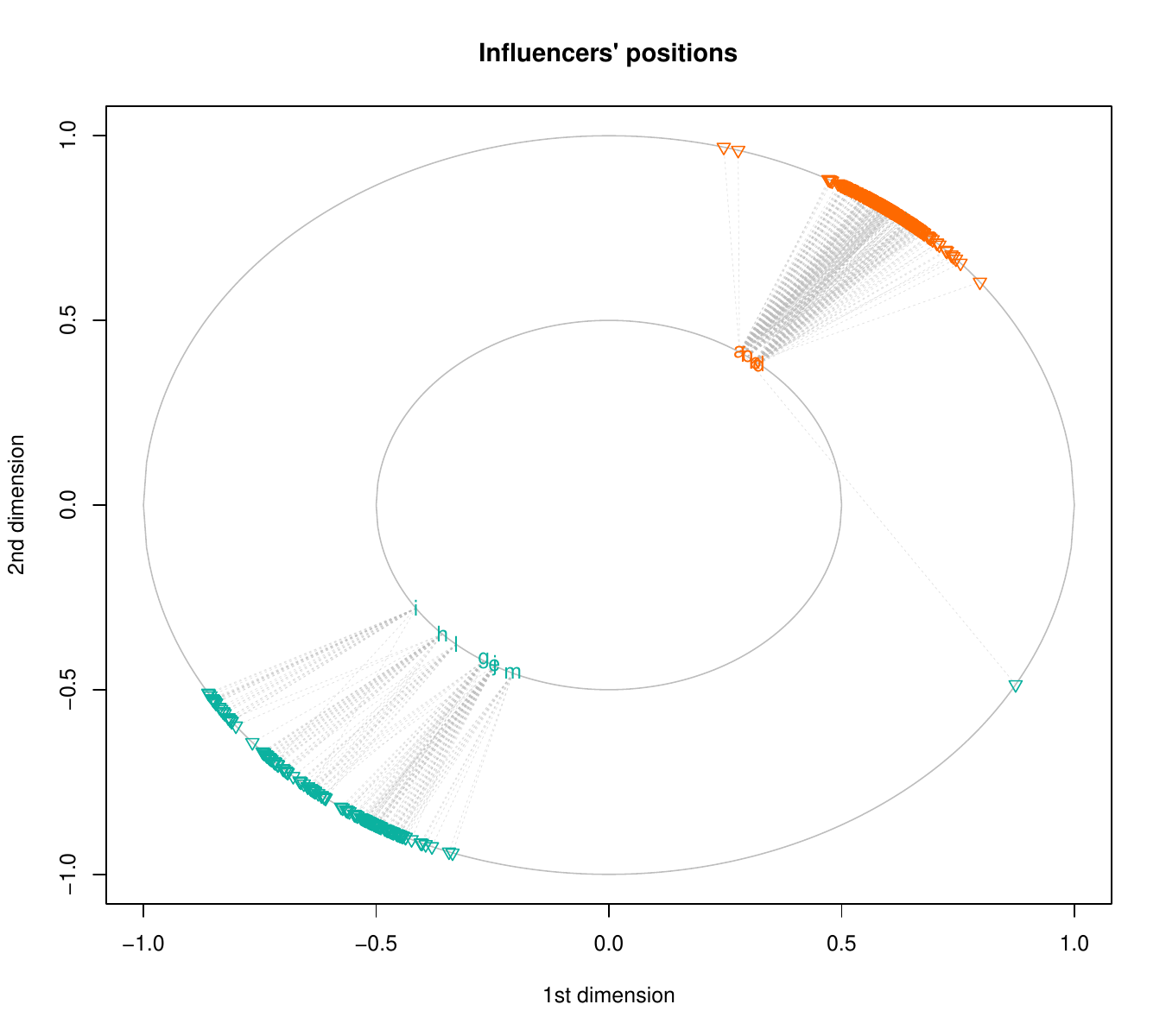}
    \caption{Positioning of individuals in the political spectrum.
    The inner circle displays the positions of the top $2\%$ of individuals with the highest influencing capacity, while the outer circle represents those influenced.
    In green, the individuals who oppose the Tax Reform, and in orange, those who support it.}
    \label{fig:rt-postura-influencia-2}
\end{figure}

Additionally, we examine the set of parameters $\thebf_1,\ldots,\thebf_N$, where $\thebf_i = \frac{1}{\Ip_i}\boldsymbol{u}_i = \frac{1}{\mathnorm{\boldsymbol{u}_i}}\boldsymbol{u}i$, for $i=1,\ldots,N$.
BY definition, these parameters are directional parameters that are located on the unit sphere in $\Real^p$, which in this case we refer to as \textit{political spectrum}.
The estimated political positions $\Esf(\thebf_i \mid \mathbf{Y}) \approx \frac{1}{B} \sum{b=1}^B \thebf_i^b$ allows us to investigate the political stances of every actor in the system.
Figure \ref{fig:rt-postura-influencia-2} displays such estimated political positions.
The inner circle shows the positions of the top $2\%$ individuals with the highest influencing capacity, whereas the outer circle represents those individuals being influenced.
Gray segments illustrate the observed influence relationships within the network. 
This figure enables a direct comparison between the political positions of influencers and those they influence.

Figure \ref{fig:rt-postura-influencia-2} reveals that most individuals, both influencers and those individuals being influenced, are located on opposite sides of the political spectrum. 
Similar to the latent space, the positions of individuals opposing the Tax Reform (green) are slightly more dispersed than those supporting it (orange).
On the other hand, the individuals influenced by the top 2\% most influential users do not have positions in other regions of the political spectrum. Consequently, we conclude that individuals whose positions are not close to the two majorities are not influenced by those with the greatest capacity to do so.
Finally, the influencers exhibit positions that are very similar to those of their followers: Neither more moderate nor more extreme.

\subsection{Goodness-of-fit}\label{sec:gof}

\begin{figure}[!htb]
    \centering
    \includegraphics[width = 0.95\textwidth]{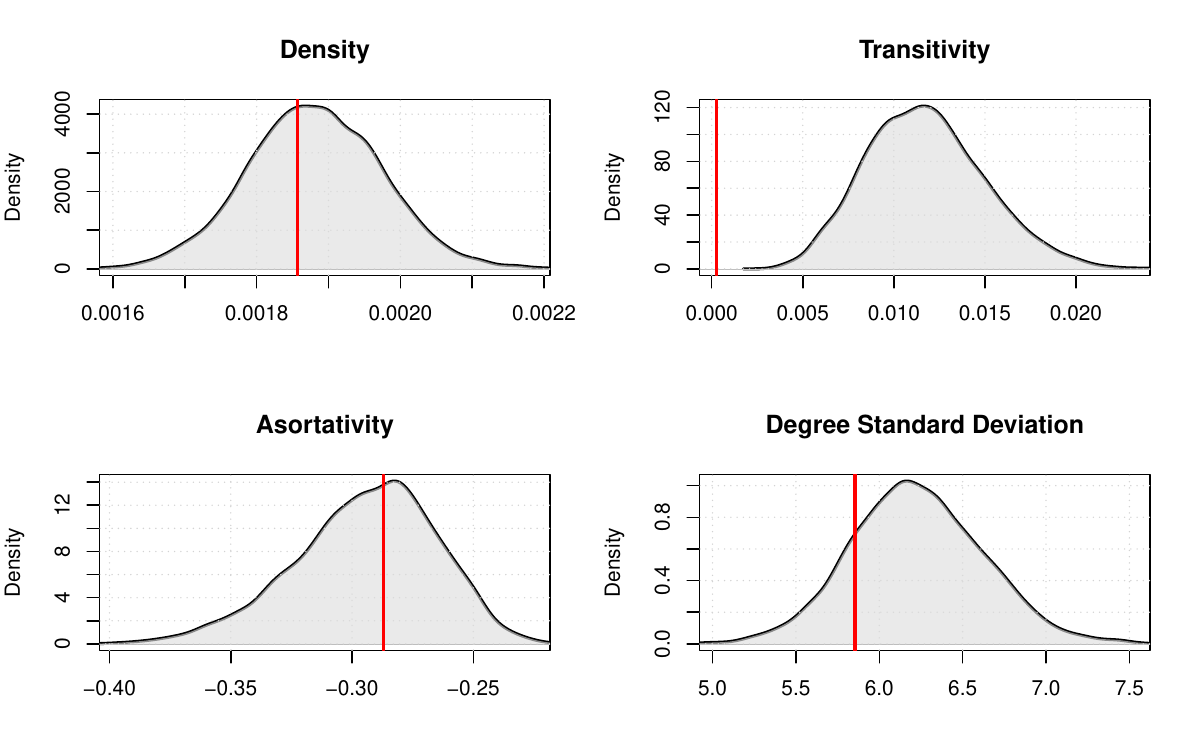}
    \caption{Kernel estimation of the posterior predictive distribution for selected network statistics. Observed values are marked in red.}
    \label{fig:rt-gof}
\end{figure}

Here, we evaluate the model's capacity to capture appropriately the network's topological structure. 
Following \cite{gelman2014bayesian}, we compare the observed value of key network statistics with the corresponding posterior predictive distribution based on the fitted model. 
We approximate this distribution from a sequence of synthetic networks generated from the samples of the posterior distribution.
The monitored statistics include density, transitivity, assortativity, and the standard deviation of the degree.
The results are shown in Figure \ref{fig:rt-gof}.
The model successfully captures all the test statistics except for transitivity. However, this does not invalidate our findings, as the observed transitivity value is nearly zero, and the posterior predictive distribution reaches at most $0.025$.

Now, we also evaluate the model's power to estimate the ``true'' parameters that generate the network. 
For this purpose, we fix the model parameters $\Op_i$ and $\ubf_i$ to the corresponding posterior means obtained in the case study, for $i=1,\ldots,N$ with $N=634$. 
Then, we randomly generate a network based on these predetermined values using the sampling distribution \eqref{eq:likelihoodhoff}.
Next, we subsequently fit the model using the synthetic network as input, following the same procedure as in the case study (i.e., $p=2$ and $a_\omega=b_\omega=a_\sigma=b_\sigma=1$).
Next, we obtain $95\%$ credibility intervals for each $\Op_i$ and each $\ubf_i$, based on $B=5000$ effective samples from the posterior distribution.
Thus, the proportion of times a parameter falls within its corresponding credibility interval quantifies the ability of the model to estimate such parameter accurately.
The results indicate that 95.11\% of the credible intervals for $\Op_i$ contain the true parameter values, while 99.53\% of the credible intervals for $\ubf_i$ do so as well.
These findings strongly support that our proposed methodology can accurately estimate the parameters of the model.

\section{Simulation for the diffusion model}\label{sec:difsimulation}

In this section, we conduct an exhaustive simulation study based on the model proposed in Section \ref{sec:difusion}. 
Twelve distinct simulation scenarios were each repeated four times, resulting in a comprehensive total of 48 experiments.
After each simulation, the following characteristics are calculated: Total diffusion time (cumulative time of each cascade jump) and diffusion reach (proportion of individuals in state S or R).
Finally, the results are analyzed using ANOVA models.

\subsection{Simulation scenarios}\label{sec:scenarios}

The simulation scenarios are defined by varying the characteristics of the individuals, the structure of the influence network, and the selection of initiators (individuals who are not in state I at the beginning of the cascade). 

For setting the individuals' influencing capacity, we consider the following cases: 
\begin{enumerate} 
    \item All individuals have the same influencing capacity $\Op_i = 4 - k^*$, where $k^*$ is a scalar, ensuring that the expected average degree of the simulated network is 10 (this value is calculated numerically).
    \item The influencing capacity of individuals follows a left-shifted Gamma distribution $k^*$ units, i.e., $\Op_i + k^* \sim \textsf{Gamma}(3,3/4)$. 
    The Gamma distribution is chosen to represent scenarios in which a large proportion of individuals have low influencing capacity. In contrast, a smaller proportion is highly influential (mean equal to 4.000, and 75th percentile equal to 5.227). 
    Any other positively skewed distribution can also be used. 
\end{enumerate}

Then, for setting the individuals' distribution of susceptibility, we consider the following cases:
 \begin{enumerate}
    \item All individuals have the same level of susceptibility $\Ip_i = 2$. 
    \item Individual susceptibility follows a Gamma distribution, i.e., $\Ip_i \sim \textsf{Gamma}(14/5,7/5)$.
    The Gamma distribution is selected to describe cases where a large proportion of individuals are not very susceptible.
    In contrast, a smaller proportion is highly susceptible (mean equal to 2.000 and 75th percentile equal to 2.628). 
    Any other positively skewed distribution can also be used. 
\end{enumerate}

Next, for setting the influence network structure and the selection of initiators, we consider the following cases: 
\begin{enumerate} 
    \item Influence network with partition modularity of two communities less than 0.001 (network with poorly defined communities). 
    \begin{enumerate}
        \item Initiators are chosen randomly, one in state S and the other in state R. 
        \item Initiators are the two individuals with the highest influencing capacity $\Op_i$, the first one in state S and the second one in state R (this case applies only if $\Op_i$ is not constant). 
    \end{enumerate} 
    \item Influence network with partition modularity of two communities greater than 0.05 (network with well-defined communities). 
    \begin{enumerate}
        \item Within each of the two communities, an initiator is chosen randomly. One initiator starts in state S and the other in state R.
        \item Initiators are the individuals with the highest influencing capacity $\Op_i$ in each community. 
        One initiator starts in state S and the other in state R (this applies only if $\Op_i$ is not constant). 
    \end{enumerate}
\end{enumerate}

Finally, in order to simulate the networks according to the defined characteristics, we generate the values of $\tau_{i,j}$, for $i,j=1,\ldots,N$, such that the resulting network has the desired level of clustering according to a given scenario. 
The values of $\tau_{i,j}$ are simulated as follows:
\begin{equation*}
    \tau_{i,j} \sim \left\{\begin{array}{cc}
        1 -  2\Betasf\left(1 - \kappa, 1\right) & \text{ si } \xi_i = \xi_j; \\
        -1 + 2\Betasf\left(1 - \kappa, 1\right) & \text{ si } \xi_i \neq \xi_j\,,
    \end{array}\right.
\end{equation*}
where $0 \leq \kappa \leq 1$ is an appropriately chosen constant and $\xi_i$ is the group to which the individual $i$ belongs, for $i=1\ldots,N$. 
It can be shown that $\covsf(\tau_{i,j},\tau_{i',j'}) = 0$ for any $i,i',j,j'=1,\ldots,N$, with $i\neq i'$ and $j\neq j'$. 
Therefore, $\lim_{\kappa\to 0}\Esf(\tau_{i,j})=0$ and $\lim_{\kappa\to 1}\Esf(\tau_{i,j}) = -1 + 2 1_{{i=j}}$, and $\CVsf(\tau_{i,j}) = \frac{2}{\kappa}\sqrt{\frac{1-\kappa}{3-\kappa}}$, for $\kappa>0$ and $i \neq j$. 
Consequently, values of $\kappa$ close to one generate networks with two communities where individuals have opposite stances but strong affinity within their groups, while values of $\kappa$ close to zero  generate networks with individuals having uniformly distributed opinions.

\subsection{Results}

Our synthetic diffusion processes are simulated in networks of size \(N=1000\). 
Based on the results, we implemented two ANOVA models to identify the factors that determine whether the simulations have a statistically significant impact on the monitored characteristics.
Since 48 simulations are conducted, this is the number of observations we use to construct the two ANOVA tables, using a significance level of \(\alpha = 5\%\). 
The first and second ANOVA models use the natural logarithm of the diffusion time and the the natural logarithm of the diffusion reach as the response variable, respectively.
It is important to note that the sampling factor (method of selecting the initiators) is nested within the modularity factor (degree of network clustering), i.e., the sampling depends on the network structure.

\begin{table}[!htb]\footnotesize
\centering
\setlength{\tabcolsep}{2.5pt} 
\begin{tabular}{lrrrr}
\hline
\multicolumn{1}{c}{\multirow{2}{*}{\textbf{\begin{tabular}[c]{@{}c@{}}Source of \\ variation\end{tabular}}}} & \multicolumn{2}{c}{\textbf{\begin{tabular}[c]{@{}c@{}} Diffusion time\end{tabular}}}                        & \multicolumn{2}{c}{\textbf{\begin{tabular}[c]{@{}c@{}} Diffusion reach\end{tabular}}}                            \\ \cline{2-5} 
\multicolumn{1}{c}{}                                                                                         & \multicolumn{1}{c}{\textbf{Coefficient}}                                      & \multicolumn{1}{c}{\textbf{$p$ value}} & \multicolumn{1}{c}{\textbf{Coefficient}}                                       & \multicolumn{1}{c}{\textbf{$p$ value}} \\ \hline
Influencing capacity                                                            & 5.861 (\textit{2})                                           & 0.0000                               & -1.06 (\textit{2})                                           & 0.0000                               \\
Susceptibility to influence                                                             & 0.819 (\textit{2})                                           & 0.0287                               & 0.396 (\textit{2})                                             & 0.0000                               \\
Modularity                                                        & -1.372 (\textit{2})                                          & 0.4997                               & -0.013 (\textit{2})                                           & 0.3816                               \\
Sampling/Modularity                                               & 0.900 (\textit{2.a}) , -1.543 (\textit{1.b})  & 0.0178                               & -0.130 (\textit{2.a}) , -0.145 (\textit{1.b})  & 0.1387                               \\ \hline
\end{tabular}
\caption{Next to the the coefficients, we place the corresponding simulation scenario or configuration, according to the numbering assigned in the Section \ref{sec:scenarios}.}
\label{tab:sims-anova}
\end{table}

Table \ref{tab:sims-anova} shows that the distribution of influencing capacity (\(\Op\)) and the distribution of susceptibility to influence (\(\Ip\)) have a significant effect on the diffusion times.
These factors assume constant values that speed up the diffusion. On the other hand, the sampling factor is also significant, and since it is nested within network modularity, the latter also turns out to be significant in the model.
Therefore, the degree of network clustering affects the diffusion time. 
Specifically, a clustered influence network significantly reduces the diffusion time.

Moreover, the distribution of influencing capacity (\(\Op\)) and the distribution of susceptibility to influence (\(\Ip\)) also impact the diffusion reach.
More precisely, having \(\Op\) constant has a positive effect, while having \(\Ip\) constant has a negative effect. 
On the other hand, we find that neither the method of selecting the diffusion initiators (sampling) nor the degree of network clustering (modularity) is significantly associated with the diffusion reach.

For both models, the underlying assumptions supporting their results were validated. The independence of observations was ensured by using different random seeds for the simulations. Neither the Shapiro-Wilk nor the Jarque-Bera test rejected the hypothesis of normality of the residuals. Additionally, Levene's test did not reject the hypothesis of homoscedasticity of the residuals in either model.

\section{Discussion}\label{sec:discusion}

This paper proposes a novel reparameterization of a latent space projection model \citep{hoff2002latent} to estimate three types of parameters for each individual in a complex network. 
In a debate context, these parameters correspond to the capacity to influencing, the susceptibility to influence, and the position in the social space. 
This modeling strategy introduces a innovative mechanism to fully characterize the diffusion of an idea as a function of the estimated latent characteristics, assuming that each individual can adopt one of the following states: Unknonw, Undecided, Support, or Reject the idea.

Additionally, this study examines an influence network on \texttt{Twitter} (now $\mathbb{X}$) regarding tweets, retweets, quotes, and comments published in Spanish about the Tax Reform in Colombia, during the 32 hours leading up to its
approval by the Colombian Congress on November 4, 2022 (Law 2277 of 2022). 
In this network, influence relationships are established through retweets, where an ideological polarization regarding the reform is observed.

The experiments associated with various diffusion scenarios are analyzed using ANOVA models. 
The results suggest that both the distribution of the capacity to influence and the distribution of susceptibility to influence significantly impact the diffusion time as well as the network's clustering coefficient. 
Additionally, these distributions also significantly affect the diffusion reach. 
Our findings also indicate that neither the method of selecting the diffusion initiators (sampling) nor the
degree of network clustering (modularity) is significantly associated with the diffusion
reach.

This work opens up several avenues for further research. 
On the one hand, various alternatives can also be considered for modeling the sociability of actors, including other latent space models(e.g., \citealt{sosa2021review}), alternative estimation methods to Monte Carlo (e.g., \citealt{blei2017variational}), and possibly a nonparametric approach to quantify the uncertainty associated with the random mechanism that produces the connections (e.g., \citealt{muller2015bayesian}). 
On the other hand, regarding the diffusion process, the proposed method could be extended to consider more states and even modify the way transitions are made. 
For this purpose, a more exhaustive simulation study would be ideal for evaluating more characteristics than those considered here.

\bibliography{references}
\bibliographystyle{apalike}

\clearpage
\appendix

\section{MCMC algorithm}

The following are the full conditional distributions of the parameters associated with the influence model:
{\footnotesize
\begin{align*}
    \psf \left( \Op_i  \mid  \text{resto} \right) & \propto \expsf \left[ -\frac{1}{2} \left(\frac{\Op_i}{\omega}\right)^2 \right] \\
    &\qquad\times\prod_{j=1 : j\neq i}^N \expitsf \left(\Op_i + ||\ubf_i||^{-1}(\ubf_i^\tr \ubf_j) \right)^{y_{i,j}} 
        \left[1-\expitsf \left(\Op_i + ||\ubf_i||^{-1} (\ubf_i^\tr \ubf_j) \right)\right]^{1-y_{i,j}} \\
    \psf \left( \ubf_i  \mid  \text{resto} \right) & \propto \expsf \left[ -\frac{1}{2} \left(\frac{||\ubf_i||}{\sigma}\right)^2 \right] \\
   &\qquad\times\prod_{j=1 : j\neq i}^N \expitsf \left(\Op_i + ||\ubf_i||^{-1}(\ubf_i^\tr \ubf_j) \right)^{y_{i,j}} 
        \left[1-\expitsf \left(\Op_i + ||\ubf_i||^{-1} (\ubf_i^\tr \ubf_j) \right)\right]^{1-y_{i,j}} \\
    \omega^2  \mid  \textsf{resto} & \sim \GammaIsf \left(a_\omega + \frac{N}{2}, b_\omega + \frac{1}{2} \sum_{i=1}^N \Op_i^2 \right)\\
    \sigma^2  \mid  \textsf{resto} & \sim \GammaIsf \left(a_\sigma + \frac{Np}{2}, b_\sigma + \frac{1}{2} \sum_{i=1}^N ||\ubf_i||^2 \right)\\
\end{align*}
}

To generate $B$ samples from the posterior distribution, an initial set of values $\mathbf{\Upsilon}^{0}$ must be established, and then, $\mathbf{\Upsilon}^{b+1}$ is generated from $\mathbf{\Upsilon}^{b}$, for $b = 1, \ldots, B$, as follows:

\begin{itemize}
    \item Sample $(\omega^2)^{(b+1)} \sim \GammaIsf \left(a_\omega + \frac{N}{2}, b_\omega + \frac{1}{2} \sum_{i=1}^N \left(\Op_i^{(b)}\right)^2\right)$
    \item For $i = 1, \ldots, N$:
    \begin{itemize}
        \item Generate a proposal $\Op_i^{(*)} \sim \Normalsf\left(\Op_i^{(b)},\delta_i^{(\Op)}\right)$, where $\delta_i^{(\Op)}$ is a tunning parameter.
        \item Compute the acceptance criterion: $$r = \frac{\psf \left( \Op_i^{(*)}  \mid  \mathbf{Y} \right)}{ \psf \left( \Op_i^{(b)}  \mid  \mathbf{Y} \right) }$$
        \item If $r > 1$, set $\Op_i^{(b+1)} = \Op_i^{(*)}$; otherwise, set
        $$\Op_i^{(b+1)} = \left\{\begin{array}{ll}
            \Op_i^{(*)} & \text{ with probability }r, \\
            \Op_i^{(b)} & \text{ with probability }1-r.
        \end{array}\right.$$
    \end{itemize}
    \item Sample $\left(\sigma^2\right)^{(b+1)} \sim \GammaIsf \left(a_\sigma + \frac{Np}{2}, b_\sigma + \frac{1}{2} \sum_{i=1}^N \mathnorm{\ubf_i^{(b)}}^2 \right)$
    \item For $i = 1, \ldots, N$:
    \begin{itemize}
        \item Generate a proposal $\ubf_i^{(*)} \sim \Normalsf_p\left(\ubf_i^{(b)},\delta_i^{(\ubf)}\Ibf\right)$, where $\delta_i^{(\ubf)}$ is a tunning parameter.
        \item Compute the acceptance criterion: $$r = \frac{\psf \left( \ubf_i^{(*)}  \mid  \mathbf{Y} \right)}{ \psf \left( \ubf_i^{(b)}  \mid  \mathbf{Y} \right) }$$
        \item If $r > 1$, set $\ubf_i^{(b+1)} = \ubf_i^{(*)}$; otherwise, set
        $$\ubf_i^{(b+1)} = \left\{\begin{array}{ll}
            \ubf_i^{(*)} & \text{ with probability }r, \\
            \ubf_i^{(b)} & \text{ with probability }1-r.
        \end{array}\right.$$
    \end{itemize}
\end{itemize}

\section{Diffusion algorithm}\label{anx:difusion}

Pseudo-code for simulating a cascade based on the model presented in Section \ref{sec:difusion}. The notation is as follows:

\begin{itemize}
    \item $V$ set of vertices.
    \item $E$ set of edges.
    \item $\tau$ matrix with entries $\tau_{i,j}$.
    \item $k$ number of consecutive influence jumps required to stop the algorithm.
    \item $jump\_times$ vector that will store the elapsed time between each jump.
    \item $.append()$ function that adds a new value to the end of the vector that calls it.
    \item $still\_stable()$ function that indicates if the number of individuals in each group has varied by less than 5\% of the total since it started occurring consecutively.
\end{itemize}

Additionally, each node $u \in V$ has the following attributes:
\begin{itemize}
    \item $u.state$ state of the individual. Initially, it is usually $I$.
    \item $u.O$ = $\Op_u$.
    \item $u.I$ = $\Ip_u$.
    \item $u.out\_neighbors$ = $\mathcal{N}^{(out)}(u)$.
\end{itemize}

\begin{algorithm}
\caption{Diffusion or cascade in an influence network}\label{alg:difusion}
\begin{tiny}
\begin{algorithmic}[2]
\Require $V$, $E$, $tau$, $k$.
\Ensure $u.state$ for $u\in V$, $jump\_times$.

\State $jump\_times := ()$
\State $n\_consecutive\_stable\_jumps := 0$
\While{$n\_consecutive\_stable\_jumps < 3N$ } \Comment{Start iterator.}
    \State // 1) Simulate time until each influence and information jump.
    \State $min\_t := \infty$
    \State $argmin\_t.u := 0$
    \State $argmin\_t.v := 0$
    
    \For{$u \in V$} \Comment{Loop through influencers or informers.}
        \If{$u.state = ``I"$}
            \State continue \Comment{Go to the next actor.}
        \EndIf
        \For{$v \in u.out\_neighbors$} \Comment{Loop through the out-neighbors of $u$.}
            \If{$v.state = ``I"$} \Comment{Information jumps}
                \State $t \gets \operatorname{Exponential}(exp(u.O))$ \Comment{Simulate $\kappa_{u,v}$.}
                \If{$t < min\_t$}
                    \State $argmin\_t.u \gets u$
                    \State $argmin\_t.v \gets v$
                    \State $min\_t \gets t$
                \EndIf
                \State continue \Comment{Go to the next $v$ out-neighbor of $u$.}
            \EndIf
            \If{$u.state = ``U"$}
                \State continue \Comment{Go to the next $v$ because $u$ can't influence, just can inform.}
            \EndIf
            \If{$v.state \neq u.state$} \Comment{Potential influence jumps.}
                \State $t \gets \operatorname{Exponential}(exp(u.O + tau(u,v)*v.I))$ \Comment{Simulate $\rho_{u,v}$.}
                \If{$t < min\_t$}
                    \State $argmin\_t.u \gets u$
                    \State $argmin\_t.v \gets v$
                    \State $min\_t \gets t$
                \EndIf
            \EndIf
            \EndFor
    \EndFor
    \State // 2) Perform the jump:
    \If{$argmin\_t.v.state = ``I"$} \Comment{Inform.}
        \State $argmin\_t.v.state \gets ``U"$
    \EndIf
    \If{$argmin\_t.v.state = ``U"$} \Comment{Influence undecided actors.}
        \State $argmin\_t.v.state \gets argmin\_t.u.state$
    \EndIf
    \If{$argmin\_t.v.state = ``R" \lor argmin\_t.v.state = ``S"$} \Comment{Influence those in $R$ or $S$.}
        \State $argmin\_t.v.state \gets ``U"$
    \EndIf
    \State // 3) Save jump time.
    \State $jump\_times.append(min\_t)$ 

    \State // 4) Compute and save the characteristics that will be monitored.
    \State // 5) Update stopping criterion:
    \If{$min\_t = \infty$}
        break
    \EndIf
    \If{$still\_stable()$}
        \State $n\_consecutive\_stable\_jumps = n\_consecutive\_stable\_jumps + 1$
    \Else
        \State $n\_consecutive\_stable\_jumps = 0$
    \EndIf
    
\EndWhile    

\end{algorithmic}
\end{tiny}
\end{algorithm}

\end{document}